\documentclass[aps,pra,twocolumn,superscriptaddress,nofootinbib]
{revtex4-1}

\usepackage{amsmath,amssymb,bm}
\usepackage{graphicx}
\usepackage[caption=false]{subfig} 

\usepackage{braket}
\usepackage{hyperref}
\hypersetup{
    colorlinks=true,
    linkcolor=blue,
    citecolor=blue,
    urlcolor=blue
}

\usepackage[utf8]{inputenc}
\usepackage[english]{babel}
\usepackage[T1]{fontenc}
\usepackage{lmodern}

\begin{document}
	\definecolor{cadmiumgreen}{rgb}{0.0, 0.42, 0.24}

\title{Distribution of Bell State Entanglement in Qubit Networks via Collision Models}

\author{Elif Yunt}
\email{elif.yunt@tau.edu.tr}
\affiliation{Department of Energy Science and Technology, Faculty of Science, Turkish-German University, 34820 Beykoz, Istanbul, Türkiye}

\author{Mert Doğan}
\email{medogan25@ku.edu.tr}
\affiliation{Department of Physics, Koç University, 34450 Sariyer, Istanbul, Türkiye}

\author{Öner Faruk Ödemiş}
\email{oenerfaruo01@zedat.fu-berlin.de}
\affiliation{Department of Physics, Freie Universität Berlin, 14195 Berlin, Germany}

\author{Özgür E. Müstecaplıoğlu}
\email{omustecap@ku.edu.tr}
\affiliation{Department of Physics, Koç University, 34450 Sariyer, Istanbul, Türkiye}
\affiliation{TÜBİTAK Research Institute for Fundamental Sciences (TBEA), 41470 Gebze, Türkiye}

\date{\today}

\begin{abstract}
	
We investigate the selective generation of Bell-pair entanglement in three-qubit networks using a standard quantum collision model. By varying the network geometry, the node contacted by the ancilla, and the interaction Hamiltonians, we identify configurations in which high pairwise concurrence and Bell-state fidelity are generated between different target qubits, including nonadjacent qubits and qubits not directly coupled to the ancilla. For comparison, we also consider the continuous-time unitary evolution of the complete ancilla–network system, sampled at the same discrete time intervals. This reference calculation clarifies how discarding each outgoing ancilla and removing its correlations from subsequent dynamics modifies the attainable entanglement and preparation time. Our results provide a proof-of-principle demonstration of geometry-dependent Bell-pair generation and routing in small qubit networks.

\end{abstract}

\maketitle



\section{Introduction}

Distributing entanglement across multiple nodes in a quantum network remains a central challenge for the realization of emerging quantum technologies. Conventional approaches often rely on entanglement swapping~\cite{Zukowski1993}, long-range unitary gates~\cite{Monz2011, Song2019}, photonic channels~\cite{Kimble2008}, or measurement-based techniques~\cite{Zwerger2012}. However, these methods frequently demand non-local control, precise long-range interactions, or real-time feedback, posing significant scalability and robustness issues, particularly in the presence of noise or architectural constraints. Addressing this challenge requires robust and controllable mechanisms for generating and distributing entanglement to selected qubits within a network.

Multipartite entanglement is a key resource for quantum information processing, spanning distributed quantum computing, secure quantum communication, and quantum-enhanced metrology~\cite{Horodecki2009, Pezze2018}. Among the various forms of multipartite entangled states, specific structures, such as the Greenberger–Horne–Zeilinger (GHZ) state, are particularly valuable. A GHZ state typically describes a maximally entangled superposition of all '0's and all '1's across multiple qubits, achieving enhanced sensitivity and Heisenberg-limited precision in quantum sensing applications~\cite{Giovannetti2006, Toth2014}. For a two-qubit system, the $|\tilde{\Phi}^+\rangle:=(|00\rangle + i |11\rangle)/\sqrt{2}$ state is a specific type of Bell state that exhibits this 'all-or-nothing' correlation with its structural resemblance to larger GHZ states. Generating and distributing these Bell states in a controllable manner to selected qubits within a quantum network, by choosing proper ancilla-system connection patterns, is a critical step towards advanced quantum applications.

An alternative framework for engineering correlations and studying open quantum system dynamics is provided by \textit{collision models}~\cite{Ciccarello2017, Scarani2002,CICCARELLO20221}. In these models, a system interacts sequentially with ancillary environmental units. This approach provides an intuitive and physically transparent description of dissipation and decoherence while retaining a clear microscopic interpretation of system-environment interactions \cite{PhysRevA.87.040103,PhysRevX.7.021003,RevModPhys.93.035008}. Originally developed to microscopically simulate Markovian dynamics, these models have been generalized to describe non-Markovian processes with environmental memory and information backflow~\cite{PhysRevA.87.040103, Lorenzo2017,PhysRevA.87.040103,PhysRevA.87.030101}, thermalization~\cite{Karevski2009}, decoherence~\cite{McCloskey2014}, and, more recently, quantum homogenization~\cite{Yosifov2025, Yosifov2024, Karpat2025}.

They have  been extended to cascaded quantum systems and quantum networks, establishing strong connections with the input-output formalism, cascaded master equations, and quantum feedback network theory \cite{PhysRevA.31.3761,PhysRevLett.70.2269,PhysRevLett.70.2273,PhysRevA.50.1792,PhysRevA.95.053838,PhysRevA.97.053811,Gough2009CMP,Gough2009IEEE,PhysRevA.78.062104,PhysRevA.81.023804,Gough2016JMP}. Owing to their conceptual simplicity and broad applicability, quantum collision models have become a versatile framework for studying quantum thermodynamics, coherence generation, quantum thermometry, nonequilibrium dynamics, and many-body quantum systems \cite{Cusumano2024NJP,Corr2025QST,Pezzutto2025QST,PhysRevLett.123.140601,PhysRevLett.123.180602,PhysRevA.104.052214,PhysRevA.105.042601,Lieb1972CMP,PhysRevLett.97.050401,PhysRevLett.96.136801,PhysRevLett.97.150404,RevModPhys.83.863}.

While collision models have demonstrated potential for quantum state engineering and entanglement generation~\cite{Ziman2005, Giovannetti2012}, their systematic use in controllably distributing specific types of multipartite entanglement within a quantum network has received limited attention. 

We investigate the selective generation and routing of Bell-pair entanglement in three-qubit networks. By varying the network geometry, the node contacted by the ancilla, and the form of the network and ancilla–system interactions, we identify configurations in which high pairwise concurrence and Bell-state fidelity are generated between different target pairs, including qubits that do not directly interact with the ancilla or with each other. The principal open-system setting is a standard fresh-ancilla collision model. For reference, we also consider the continuous-time unitary evolution of the complete ancilla–network system, sampled at the same discrete time intervals. This comparison clarifies how discarding each outgoing ancilla and removing its correlations from subsequent dynamics modifies geometry-dependent entanglement generation.

We investigate these models in small networks of three qubits arranged in open-chain and closed-loop geometries. By tuning the interaction Hamiltonians and the qubit that the ancilla collides with, we demonstrate selective generation of Bell pairs between target qubit pairs—even when those qubits are not directly coupled. This form of dynamic, coherence-mediated routing enables long-range correlations to be established in a controllable way. Our results provide a route to distributed state preparation based on minimal control assumptions. 

The remainder of this paper is organized as follows. Sec.~\ref{sec:modelNmethods} introduces the qubit network geometries and Hamiltonians under consideration, outlines the theoretical modeling framework and describes the ancilla–network interaction protocol. In Sec.~\ref{sec:results}, we present numerical results on $|\tilde{\Phi}^\pm\rangle$ type bipartite entanglement generation with a comparison between two different geometrical configurations. We conclude in Sec.~\ref{sec:conclusion} with a discussion of our results. In the Appendix A and B, we discuss two limiting cases in which the ancilla interaction reduces to a coherent local field.


\begin{figure*}[t!]
	\centering
	\subfloat[]{\includegraphics[width=0.33\textwidth]{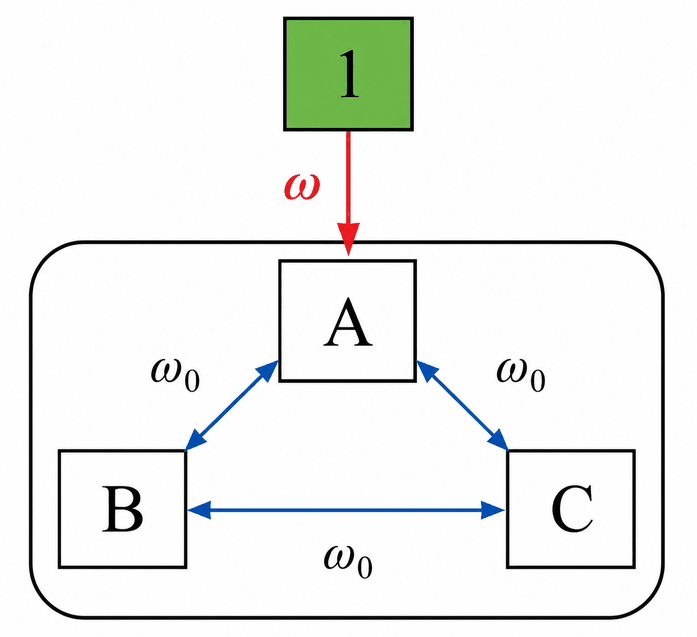}	\label{fig:closed}} \hspace{2cm}\quad 
	\subfloat[]{\includegraphics[ width=0.41\textwidth]{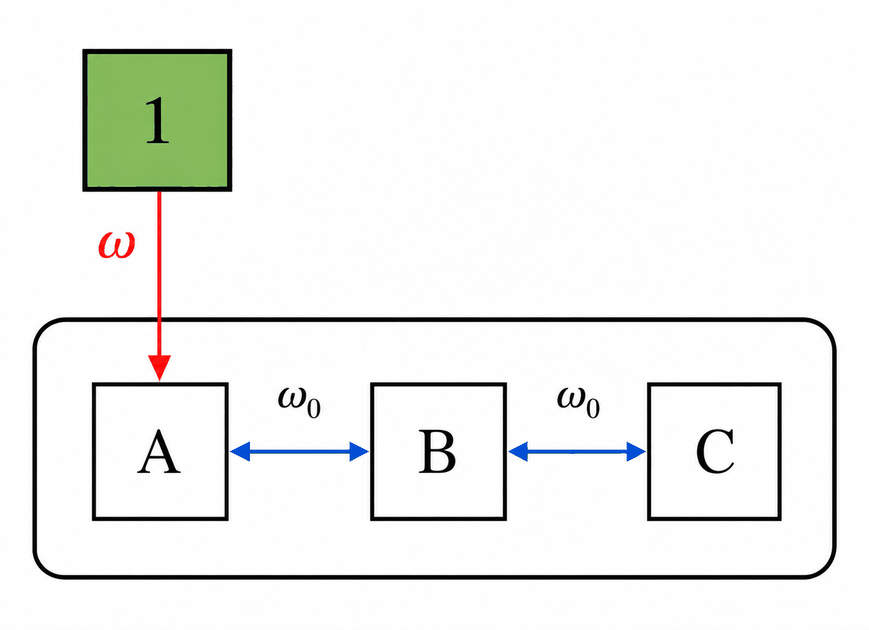}	\label{fig:edge}}

	\caption{(Color Online) Illustration of the three-qubit network configurations and the local ancilla coupling. The system consists of qubits A, B, and C, with the ancilla qubit designated as 1. (a)  Closed-loop (triangular) network, with the ancilla coupled to qubit A. (b) Open linear chain, with the ancilla coupled to the terminal qubit A. The internal network coupling strength is denoted by $\omega_0,$  and the ancilla-network coupling strength by $\omega.$}
	\label{fig:collisionconfigurations}
\end{figure*}



\section{Model and Methods}
\label{sec:modelNmethods}

We consider a network of three qubits labeled A, B, and C, interacting via a system Hamiltonian $H_\mathrm{S}$, and coupled to an ancillary qubit labeled $1$ through a tunable interaction Hamiltonian $H_\mathrm{int}$. The ancilla serves as a mediator to distribute coherence and entanglement across the network. Each interaction occurs in discrete time steps $\Delta t$, and the total evolution is treated as a sequence of system–ancilla collisions.

\subsection{Model}
\label{subsec:model}

We consider a network of three qubits, labeled A, B, and C. These qubits are arranged in either an open linear chain or a closed triangular geometry as depicted in Fig.~\ref{fig:collisionconfigurations}. The connectivity between the qubits is encoded in the adjacency matrix $\epsilon_{ij}$, where $\epsilon_{ij} = 1$ if qubits $i$ and $j$ are directly coupled and $\epsilon_{ij} = 0$ otherwise.

The system Hamiltonian governing the internal interactions among the network qubits is defined as
\begin{equation}
H_\mathrm{S} = \omega_0 \sum_{i<j} \epsilon_{ij} \, h_{ij},
\label{eq:Hsystem}
\end{equation}
where $\omega_0$ is the uniform qubit–qubit coupling strength in the system. We consider three types of pairwise interactions, one of which is selected per simulation:
\begin{itemize}
    \item XX-type coupling: $h_{ij} = \sigma^x_i \sigma^x_j$,
    \item ZZ-type coupling: $h_{ij} = \sigma^z_i \sigma^z_j$,
    \item Exchange coupling: $h_{ij} = \frac{1}{2}(\sigma^+_i \sigma^-_j + \sigma^-_i \sigma^+_j)$.
\end{itemize}
Here, $\sigma^\alpha_i$ are the Pauli operators for the $i$-th qubit, and $\sigma_i^\pm = (\sigma^x_i \pm i \sigma^y_i)/2$ are the ladder operators.

In the linear geometry, only nearest-neighbor pairs (A,B) and (B,C) are coupled, i.e., $\epsilon_{AB} = \epsilon_{BC} = 1$, $\epsilon_{AC} = 0$. In the triangular geometry, all three qubits are  pairwise coupled, $\epsilon_{ij} = 1$ for all $i \ne j$.

A fourth qubit, labeled as ancilla qubit $1$, interacts locally with a selected system qubit during each collision. The ancilla may couple to a terminal node (qubit $A$ or $C$) or to the central node (qubit $B$) in the linear configuration, or to any node in the triangular geometry.

The system–ancilla interaction is described by
\begin{equation}
H_\mathrm{int} = \omega \, h_{1k},
\label{eq:Hint_general}
\end{equation}
where $k \in \{A, B, C\}$ is the system qubit the ancilla couples to, and $\omega$ is the ancilla coupling strength. As with the system Hamiltonian, we consider three forms for $h_{1k}$
\begin{align}
h_{1k}^{(xx)} &= \sigma_1^x \sigma_k^x, \label{eq:Hint_xx}\\
h_{1k}^{(zz)} &= \sigma_1^z \sigma_k^z, \label{eq:Hint_zz}\\
h_{1k}^{(\mathrm{ex})} &= \frac{1}{2} \left( \sigma_1^+ \sigma_k^- + \sigma_1^- \sigma_k^+ \right). \label{eq:Hint_dip}
\end{align}

Each system–ancilla interaction lasts for a finite time interval $\Delta t$ and evolves unitarily under the total Hamiltonian
\begin{equation}
H = H_\mathrm{S} + H_\mathrm{int}, \quad U = e^{-i H \Delta t}.
\end{equation}
This evolution is followed by ancilla tracing and network state update as described in the following subsections.


\subsection{Collision Model}
\label{subsec:collision}

In the standard collision model, the environment is represented by a sequence of independent ancilla qubits, each prepared in the same initial state $\rho_1^{in}.$ During the $n$th collision, a fresh ancilla
interacts with the network $S = \text{ABC}$ for a duration $\Delta t$ under
\begin{equation}
U=e^{-i(H_\mathrm{S} + H_\mathrm{int})\Delta t}
	\label{eq:unitaryevolution}
\end{equation}
The outgoing ancilla is then discarded, and the network state is updated according to
\begin{equation}
\rho_\mathrm{S}^{(n+1)}= \mathrm{Tr}_1 \left[U\left(\rho_1^{\mathrm{in}}\otimes \rho_\mathrm{S}^{(n)}\right)U^\dagger\right]	
	\label{eq:rhoS}
\end{equation}
A new ancilla, independently prepared in  $\rho_1^{in}=\ket{\Psi_1}\bra{\Psi_1},$ is introduced at the next step. Consequently, correlations generated between the network and an outgoing ancilla do not contribute to the subsequent collisions. The ancilla reset ensures identical interaction conditions throughout the dynamics.

This process is iterated for multiple collision steps. Replacing the outgoing ancilla by an independently prepared ancilla restores the same input state at every step and leads to repeated application of an identical completely positive trace-preserving map. Since each collision involves an independent ancilla prepared in the same initial state and no ancilla--ancilla correlations are introduced, the resulting dynamics is Markovian. We note that by tracing out the ancilla, information is irreversibly transferred to the environment, effectively introducing a decoherence channel into the network.


\subsection{Continuous-time Unitary Evolution}
\label{subsec:continuous}

For comparison, we also consider the continuous-time unitary evolution of the complete ancilla-network state. In this calculation, the ancilla is initialized once and remains part of the joint four-qubit system throughout the evolution. No reset, state replacement, or decorrelation operation is applied during propagation:

\begin{equation}
	\rho_{1\mathrm{S}}(t)= e^{-i H t}\rho_{1\mathrm{S}}(0)e^{i H t},  \quad H = H_\mathrm{S} + H_\mathrm{int}.
	\label{eq:rho1S}
\end{equation}

To compare the resulting dynamics with the discrete collision protocol, observables are evaluated at the sampling times 
$$ t_n=n\Delta t .$$
At the sampled times,
\begin{equation}
	\rho_{1\mathrm{S}}(t_n)= U^n\rho_{1\mathrm{S}}(0)(U^\dagger)^n, \quad  U=e^{-iH\Delta t}.
	\label{eq:rho1Ssampled}
\end{equation}
Partial traces are used only to obtain reduced states for the calculation of concurrence and fidelity. They are not used to update the state propagated at later times.

\subsection{Characterizing Bipartite Entanglement}

After each interaction step with the ancilla, we analyze and quantify the pairwise entanglement between the network qubits using concurrence, originally introduced in~\cite{Bennett1996,Wooters1998}. It takes values between 0 and 1, with $C=0$ corresponding to a separable state and $C=1$ to a maximally entangled state. For an arbitrary two-qubit mixed state $\rho_{\alpha\beta}$, the concurrence is defined as~\cite{PhysRevLett.78.5022}
\begin{align}
	C_{\alpha\beta}=\max\left(0,\lambda_4-\lambda_3-\lambda_2-\lambda_1\right),
\end{align}
where $\lambda_4>\lambda_3>\lambda_2>\lambda_1$ are the square roots of the eigenvalues of $\rho\tilde{\rho}$, arranged in descending order. Here,
	\begin{equation*}
		\tilde{\rho}=(\sigma_y\otimes\sigma_y)\rho^*(\sigma_y\otimes\sigma_y)
	\end{equation*}
is the spin-flipped density matrix~\cite{Wootters2001}.

To identify the Bell-like states obtained at the concurrence maxima, we evaluate the fidelity between the reduced two-qubit state and the target Bell states. We focus specifically on the following two maximally entangled bipartite states~\cite{Bell1964,Bennett1993}:
\begin{equation}
	\ket{\tilde{\Phi}^\pm}=\frac{1}{\sqrt{2}}
	(\ket{00}\pm i\ket{11}).
	\label{eq:BStatesphistarpm}
\end{equation}

The fidelity of the target pure state $\ket{\Phi}$ with respect to the reduced two-qubit state $\rho_{\alpha\beta}^{(n)}$ after the $n$th collision is defined as
\begin{equation*}
	(F_{\Phi})_{\alpha\beta}^{(n)}=\bra{\Phi}	\rho_{\alpha\beta}^{(n)}\ket{\Phi},
\end{equation*}
where $0\leq F_{\Phi}\leq1$. A value of $F_{\Phi}=1$ indicates that the reduced two-qubit state perfectly matches the target Bell state, whereas $F_{\Phi}=0$ implies that the two states are orthogonal.

For a network-qubit pair $\alpha, \beta \in \{A, B, C\},$ the reduced state is
$$ \rho_{\alpha\beta}=\mathrm{Tr}_{\overline{\alpha\beta}}\rho,$$
where the trace is taken over the ancilla, when present, and over the unused network qubit. All concurrences reported below refer to pairs of network qubits.


\section{Results and Discussion}\label{sec:results}

In this section, we present and discuss our results related to the continuous-time unitary evolution and the standard collision model in two subsequent sections. Pairwise entanglement between the network qubits is quantified using concurrence.

For fixed and calibrated Hamiltonian parameters, the target sampling or collision index is predetermined from the calculated concurrence and Bell-state fidelity curves. We denote by $n_*$ the earliest index at which the desired Bell-state fidelity reaches its selected target value, or, when discussing the global maximum within the observation interval, the index at which the fidelity is maximal. The corresponding preparation time is $t_*= n_*\Delta t.$ Thus, the protocol requires predetermined switching or readout at $t_*,$ rather than state tomography after each interaction.  


\subsection{Continuous-time Unitary Evolution}\label{sec:CUE}

We first investigate the dynamics under a continuous time unitary evolution, where the whole ancilla-network system evolves unitarily with time. There is no reset of the ancilla state. This scenario represents a closed-system model in the sense that no coherence is continuously injected from an external source to the network, and the ancilla's coherence can evolve and potentially degrade. Despite this, the evolution of concurrence within the qubit network is not necessarily periodic (see Appendix B), but it also does not exhibit continuous damping, reflecting the unitary nature of each interaction step.

First, we investigate closed-loop configuration depicted in Fig.~\ref{fig:closed}. The qubits A, B and C in the network interact via the Hamiltonian
\begin{equation}
	 H_\mathrm{S} =  \omega_0(\sigma_\text{A}^x \sigma_\text{B}^x+\sigma_\text{B}^x \sigma_\text{C}^x+\sigma_\text{C}^x \sigma_\text{A}^x),
	 \label{eq:xxclosedH}	
 \end{equation}
where all qubits are pairwise connected through a XX-type coupling. We set $\hbar=\omega_0=1,$ all throughout the manuscript so that all relevant quantities like energy and time are scaled in terms of $\omega_0.$ The ancilla interacts with the qubit A through a ZZ-type interaction

\begin{equation}
	H_\mathrm{int}  = \omega\sigma^z_1 \sigma^z_\text{A}
\label{eq:zzA}	
\end{equation}
This combination is particularly interesting because the XX-interaction within the network already creates pairwise entanglement starting out from the following initial state 
 \begin{equation}
 	\ket{\Psi}_\mathrm{ABC} = \ket{000}_\mathrm{ABC} .
 	\label{eq:initialnetworkstate}	
 \end{equation}

\begin{figure}[htb!]
    \centering
    \includegraphics[width=\columnwidth]{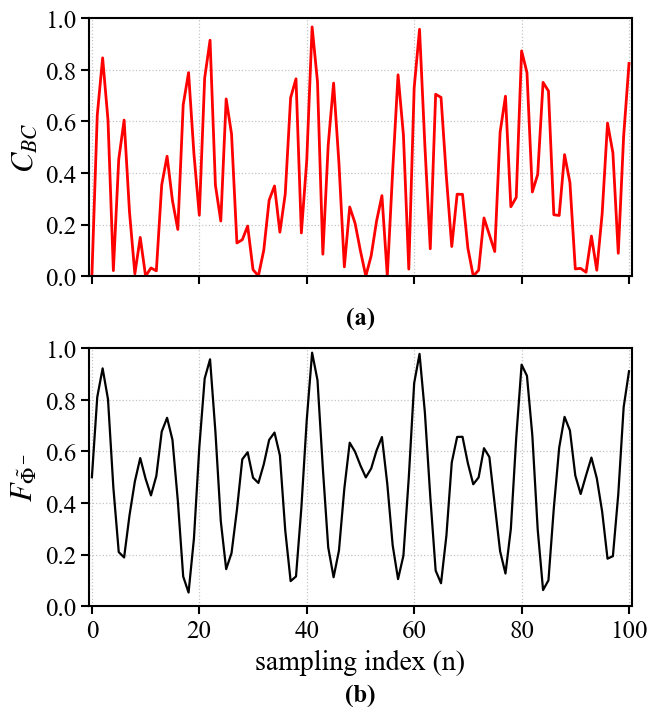}
    \caption{(Color online) (a) The concurrence $C_\text{BC}$ (red) between qubits B and C for the closed-loop triangular network (Fig.~\ref{fig:closed}) under continuous-time unitary evolution of the complete ancilla-network state, sampled at $t_n=n\Delta t.$ The Hamiltonians are $H_\mathrm{S}=\omega_0(\sigma_\text{A}^x \sigma_\text{B}^x+\sigma_\text{B}^x \sigma_\text{C}^x+\sigma_\text{C}^x \sigma_\text{A}^x), \quad H_\mathrm{int}  = \omega\sigma^z_1 \sigma^z_\text{A}.$ The  parameters are $\omega_0 = 1, \omega = 5, \Delta t = 0.4.$ The initial state is $\ket{+}_1\ket{000}_{\mathrm{ABC}}.$ (b) Fidelity with the target Bell state $\ket{\tilde{\Phi}^-}.$ The largest concurrence occurs at $n=41,$ where $C_\text{BC} \simeq 0.966$ and $F_{\tilde{\Phi}^-}\simeq0.983.$}
    \label{fig:fig2}
\end{figure}

 We work in the strong regime, where the ancilla-system coupling $\omega$ is larger than the qubit network couplings $\omega_0.$ In the absence of the ancilla interaction, the symmetric triangular XX network generates partial
 pairwise entanglement from the initial state Eq.~(\ref{eq:initialnetworkstate}), with the concurrence between qubits B and C  reaching a maximum
 value of 1/3. As the ancilla-network interaction is activated, the entanglement between qubits B and C is enhanced, reaching near-maximally entangled Bell-like states over time, as shown in Fig.~\ref{fig:fig2}. 
 
The ancilla is initialized in
 \begin{equation}
\ket{+}_1=\frac{1}{\sqrt{2}}(\ket{0}+\ket{1}),
 	\label{eq:initialancillastate}	
\end{equation}
 which assigns equal populations to the two eigenvalue sectors of
 $\sigma^z_1.$ The ancilla–network interac-
 tion consequently generates two conditional network evolutions corresponding to effective local $\pm \sigma^z_\text{A}. $ Together with the internal XX coupling, these conditional dynamics enhance the concurrence between qubits B and C.
 
 Although there is an average initial concurrence between qubits A and C (as well as between qubits A and B), it is not enhanced by the repeated interactions, so we do not include their plots in our discussion.
 
 The entanglement between B and C displays an oscillatory behavior compatible with the unitary evolution, peaking at a concurrence value of $C_\text{BC}(n=41)= 0.966$ at $n=41$, where $n$ here designates the interaction(collision) step with the ancilla. In this simulation, as well as in all the others, the time step between interactions is $\Delta t=0.4.$ The result of this simulation demonstrates that even indirectly affected qubits can achieve maximal entanglement.  We note that the same process of maximal entanglement generation can also be facilitated through an exchange-type ancilla-system interaction with the same coupling strength ($\omega=5$) for the same XX-type network Hamiltonian.
 
 In the absence of the ancilla interaction, the symmetric triangular XX network generates partial pairwise entanglement from the initial state $\ket{000},$ with a maximum concurrence $C_\text{BC}=1/3.$ Under the joint ancilla-network evolution, the concurrence is enhanced to $C_\text{BC}\simeq 0.966$ at the sampling index $n=41.$ 
 
 Next we consider the linear chain configuration given in Fig.~\ref{fig:edge} to explore the effect of geometry in our entanglement generation scheme. The network qubits again interact through an XX-type interaction, but the Hamiltonian is slightly modified from Eq.~(\ref{eq:xxclosedH}) since qubits A and C do not directly interact. Consequently, the network Hamiltonian is 
\begin{equation} 
 H_\mathrm{S} = \omega_0(\sigma_\text{A}^x \sigma_\text{B}^x+\sigma_\text{B}^x \sigma_\text{C}^x),
 \label{eq:xxopenH}
 \end{equation}
 and the ancilla interacts with the network via qubit A through the same ZZ-interaction given in Eq.~(\ref{eq:zzA}). We note that because of the symmetry of the chain, interaction of the ancilla with the network via qubit C is equivalent to interaction via qubit A.  The network starts out in the state Eq.~(\ref{eq:initialnetworkstate}) and the ancilla contains maximal initial coherence prepared in the state  Eq.~(\ref{eq:initialancillastate}).

\begin{figure*}[t!]
	\centering
	\subfloat[]{\includegraphics[width=0.48\textwidth]{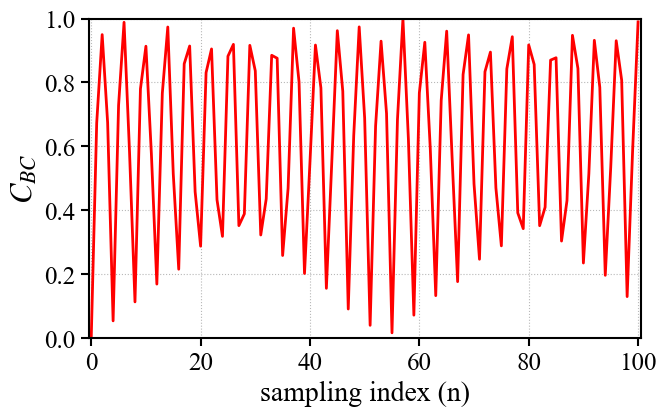}\label{fig:fig3a}}%
	\quad 
	\subfloat[]{\includegraphics[width=0.48\textwidth]{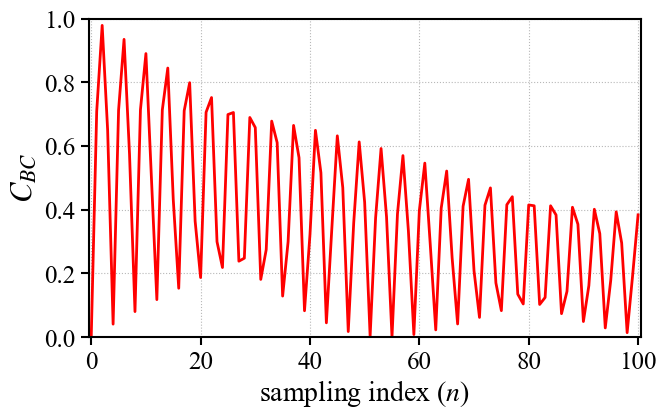}\label{fig:fig3b}}%
	
	\caption{(Color Online) Concurrence $C_\text{BC}$ (red) between qubits B and C for the open linear chain configuration (Fig.~\ref{fig:edge}), with the ancilla coupled to qubit A. The network and interaction Hamiltonians are $H_\mathrm{S} = \omega_0(\sigma_\text{A}^x \sigma_\text{B}^x+\sigma_\text{B}^x \sigma_\text{C}^x)$ and $H_\mathrm{int}  = \omega\sigma^z_1 \sigma^z_\text{A}.$ The parameters are $\omega_0 = 1, \Delta t = 0.4,$ and the initial state is $\ket{+}_1\ket{000}_{\mathrm{ABC}}.$ (a) Continuous-time unitary evolution sampled at $t_n=n\Delta t,$ with $\omega=5.$ The highest concurrence value occurs at $n=57,$ where $C_\text{BC} \simeq 0.999$ and $F_{\tilde{\Phi}^-}\simeq0.999.$(b) Standard fresh-ancilla collision model with $\omega=20.$ The largest concurrence occurs at $n=2,$ where $C_\text{BC} \simeq 0.989$ and $F_{\tilde{\Phi}^-}\simeq0.994.$}
	\label{fig:linearxxzzall}
\end{figure*}

 All concurrences other than between qubits B and C also vanish in this case, so we present in Fig.~\ref{fig:fig3a} only the time evolution of the concurrence $C_\text{BC}.$  For the linear chain geometry, the continuous-time unitary evolution reaches a near-maximal concurrence at $n=57$ with a value $C_\text{BC}(n=57)=0.998.$ Compared to the closed-loop, it takes more time in the open linear chain configuration for the network to reach the peak concurrence value between qubits B and C, but here we still observe the oscillatory concurrence evolution. 

In the next subsection, we describe the nature of these high concurrence states by calculating the fidelities of the Bell states $\ket{\tilde{\Phi}^\pm}$ given in Eq.~(\ref{eq:BStatesphistarpm}) at the maximal points of the plots.


\subsection{Characterizing High-Concurrence Bell States}\label{sec:CMES}

In this subsection, we use the concurrence measure in order to quantify the bipartite entanglement between the qubits in the network and analyze the states at the concurrence maxima. 

The highest concurrence state between qubits B and C in Fig.~\ref{fig:fig2} corresponding to the closed network configuration occurs at $n=41$ and is $\ket{\tilde{\Phi}^-}$ (given in Eq.~(\ref{eq:BStatesphistarpm}))with a fidelity value of 0.983. The state remains the same at all peaks, although naturally the concurences as well as the corresponding fidelities are lower.

The near-maximally entangled state between qubits B and C in the linear chain configuration turns out also to be $\ket{\tilde{\Phi}^-},$ but this time with a very high fidelity value equal to 0.999. This state corresponds to the peak at $n=57$ in Fig.~\ref{fig:fig3a}. Interestingly when we inspect the fidelity of the next highest concurrence peak, we find out that the achieved state is $\ket{\tilde{\Phi}^+},$ which is orthogonal to $\ket{\tilde{\Phi}^-}.$ It turns out that at every peak in Fig.~\ref{fig:fig3a}, the near-unit concurrence state changes between $\ket{\tilde{\Phi}^-}$ and  $\ket{\tilde{\Phi}^+}$ states. 

\begin{figure}[htb!]
	\centering
	\includegraphics[width=\columnwidth]{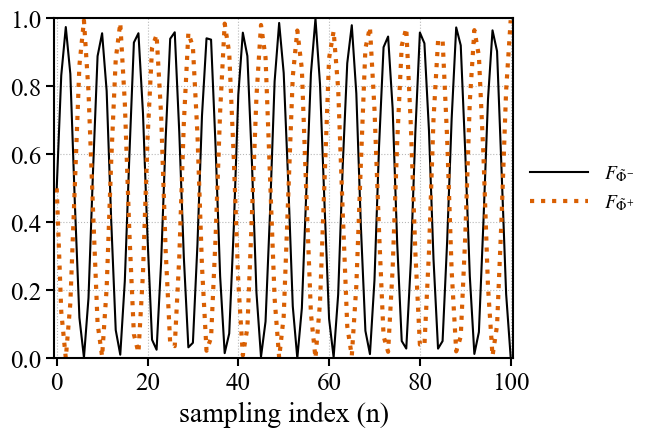}
	\caption{(Color online) Fidelities of the reduced BC state with $ \ket{\tilde{\Phi}^\pm}=\frac{1}{\sqrt{2}}
		(\ket{00}\pm i\ket{11})$ for the continuous-time unitary evolution of the linear-chain configuration shown in Fig.~\ref{fig:fig3a}. Successive high-concurrence peaks alternate between the two orthogonal Bell states. }
	\label{fig:fig3aFid}
\end{figure}

The closed-loop triangular geometry produces the same Bell-state family at the sampled
concurrence maxima (Fig.~\ref{fig:fig2}), whereas the open linear chain alternates between two orthogonal Bell states (Fig.~\ref{fig:fig3aFid}).
The triangular geometry also reaches its first high-concurrence peak at an earlier sampling time.
Within the present ideal model, it therefore offers a simpler timing and target-state selection
procedure.

In the next section, we discuss our results related to the collision model with a comparison to the continuous-time unitary evolution.

\subsection{Collision Model}\label{sec:CM}
In this section, we present our standard collision model results for the closed-loop and open linear chain geometries. In the collision model, where the ancilla qubit is reset after each interaction, we observe that in general the concurrences decay with time (see Fig.~\ref{fig:fig3b}) as expected due to the continuous interaction with a fresh environment. However, for limiting local drive cases, discussed in the Appendix A and B, decay does not take place so that continuous-time evolution and collision model results overlap (Fig.~\ref{fig:fig5} and Fig.~\ref{fig:fig6}). 

We continue the discussion of entangled state generation via the collision model by using the linear chain configuration of Fig.~\ref{fig:edge} taking into account the same network and ancilla interactions leading to Fig.~\ref{fig:fig3a}, as well as the same network and ancilla initial states given by Eq.~(\ref{eq:initialnetworkstate}) and Eq.~(\ref{eq:initialancillastate}), respectively.

We observe that in the collision model, the network-ancilla coupling $\omega$ plays a crucial role in attaining the near-maximal concurrence. As the ancilla is reset at every interaction step, the concurrence between qubits B and C decay with time as shown in  Fig.~\ref{fig:fig3b}. Near-maximal value of concurrence is achieved at a very early collision time with $C_\text{BC}(n=2)=0.989,$ and is only possible when the ancilla-system coupling is higher ($\omega=20$).

The nature of the generated state is not affected by the ancilla-reset, so that at the highest concurrence value, $\ket{\tilde{\Phi}^-}$ is achieved (with the fidelity 0.994) and at each descending peak, the state alternates between $\ket{\tilde{\Phi}^-}$ and its orthogonal counterpart $\ket{\tilde{\Phi}^+}.$

For the closed-loop configuration of Fig.~\ref{fig:fig2}, similarly, there is a rapid decay of concurrence between qubits B and C in the collision model. The initial partial entanglement generated by the XX-type interaction is enhanced briefly by ZZ-type ancilla-network interaction before it decays through damped oscillations toward zero. The enhancement in the concurrence is maximal only when the ancilla-system coupling $\omega$ is higher. With such an ancilla coupling strength $\omega=20$, the achieved state is $\ket{\tilde{\Phi}^-}$ corresponding to a peak concurrence value of $C_\text{BC}(n=2)=0.978,$ with fidelity 0.989. We conclude that despite the decay, standard collision model results in the same Bell-state family at the concurrence maxima with the continuous-time unitary evolution.

\section{Conclusion}\label{sec:conclusion}

We have studied the selective generation of Bell-pair entanglement in three-qubit networks under two clearly distinguished dynamical settings: a standard fresh-ancilla collision model and the continuous-time unitary evolution of the complete ancilla–network system. By changing the network geometry, ancilla contact node, and interaction type, different network-qubit pairs can be driven to high concurrence and Bell-state fidelity, including pairs that do not directly interact with the ancilla or with each other.

Our results demonstrate that entanglement can be controllably generated between qubits further away in the network from the contact point with the environment. We show additionally how this is also possible between non-adjacent nodes by selecting appropriate network geometries, interaction Hamiltonians and initial ancilla states.

We examined the effect of geometry under continuous-time unitary evolution by comparing the closed-loop triangular and open linear-chain configurations. The triangular geometry reaches its
first high-concurrence peak earlier and produces the same Bell-state family at the sampled maxima, whereas the linear chain alternates between two orthogonal Bell states. 

The continuous-time calculation serves as a closed-system reference for identifying the effect of discarding and replacing the ancilla in the collision protocol. Our results should be regarded as a proof-of-principle demonstration for small networks; no claim of scalability is made in the present work.

\section*{Acknowledgements}
We acknowledge support by the Scientific and Technological Research Council of Turkey (T\"{U}B{\.I}TAK), Grant No.~(125F164). 

\section*{Appendix A: Entanglement between qubits A and C due to Local z-field on Qubit B }\label{sec:appA}
\begin{figure}[t!]
	\includegraphics[width=0.41\textwidth]{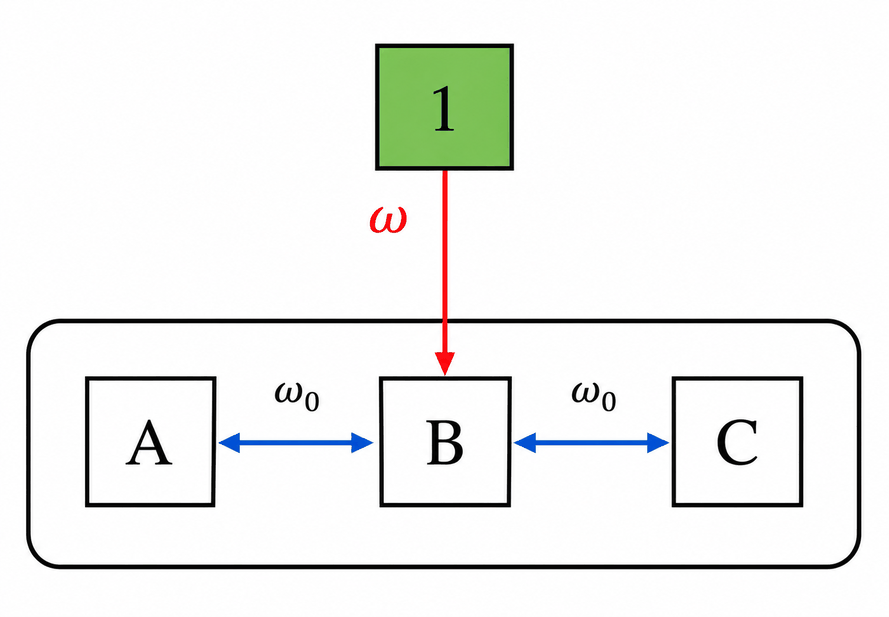}	
	\caption{(Color Online) Illustration of the three-qubit network, with the ancilla coupled to qubit B, which is the middle node in the network. The internal network coupling strength is denoted by $\omega_0,$  and the ancilla-network coupling strength by $\omega.$ }
	\label{fig:middle}
\end{figure}

\begin{figure}[htb!]
	\centering
	\includegraphics[width=\columnwidth]{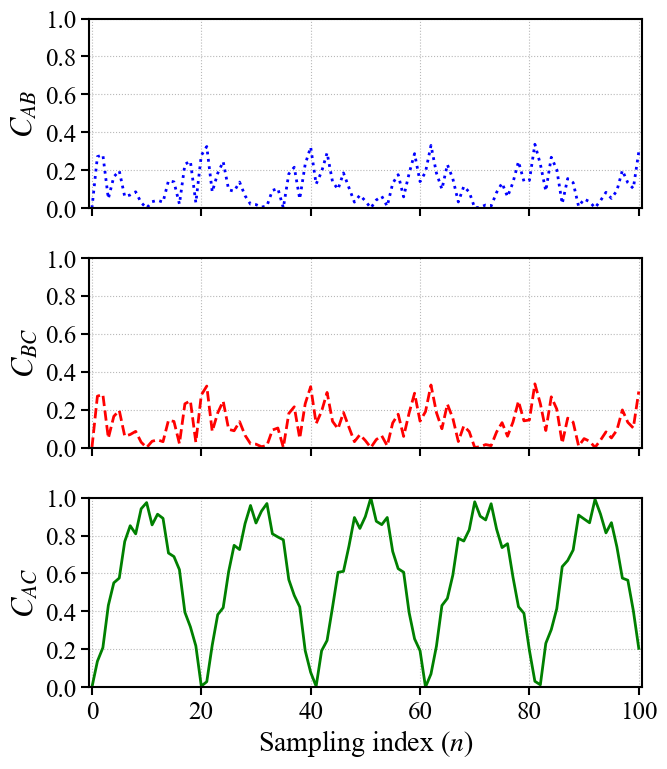}
	\caption{(Color online) Time evolution of the concurrence $C_\text{AB}(t)$ (blue, dotted), $C_\text{BC}(t)$ (red, dashed), and $C_\text{AC}(t)$ (green) for the linear chain configuration (Fig.~\ref{fig:middle}), where the ancilla interacts with the middle qubit B. The Hamiltonians are $H_\mathrm{S} =  \omega_0(\sigma_\text{A}^x \sigma_\text{B}^x+\sigma_\text{B}^x \sigma_\text{C}^x)$ and $H_\mathrm{int}  = \omega\sigma^z_1 \sigma^z_\text{B}$. The parameters are $\omega_0 = 1, \omega = 5, \Delta t = 0.4$. The initial state is $\ket{1}_1\ket{000}_{\mathrm{ABC}}.$ The ancilla is initialized in an eigenstate of $\sigma^z_1.$ Consequently, the ancilla interaction is equivalent in this sector to a coherent local z-field on qubit B. The continuous-time and fresh-ancilla collision calculations therefore produce identical network dynamics for this case.}
	\label{fig:fig5}
\end{figure}

The evolution of bipartite entanglement between each pair of qubits for the open linear chain configuration, where the ancilla interacts with the middle qubit B as depicted in Fig.~\ref{fig:middle}, is demonstrated in Fig.~\ref{fig:fig5}.
The ABC network is initially in the state given by Eq.~(\ref{eq:initialnetworkstate}). The network qubits interact via the system Hamiltonian 
$$H_\mathrm{S} =  \omega_0(\sigma_\text{A}^x \sigma_\text{B}^x+\sigma_\text{B}^x \sigma_\text{C}^x),$$ 
which already entangles them partially; however not maximally. The  ancilla is initialized, differently than all other cases discussed in the manuscript, in a non-coherent pure state given by
$$\ket{\Psi_1}=\ket{1}.$$
It interacts repeatedly with the middle qubit B via a ZZ-type interaction given by
$$ H_\mathrm{int}  = \omega\sigma^z_1 \sigma^z_\text{B}.$$ 

Because the ancilla is initialized in an eigenstate of $\sigma^z_1$, the interaction is equivalent to a static local z-field acting on qubit B. Together with the internal XX couplings, this local field generates
near-maximal entanglement between the nonadjacent qubits A and C (Fig.~\ref{fig:fig5} (green)).

 The state with the highest concurrence value in Fig.~\ref{fig:fig5} (green) emerges at $n=51$ with a concurrence value $C_\text{AC}(n=51)=0.999.$  

Other qubit pairs (AB and BC) do not achieve at any instant during the simulation near-maximal concurrence values. Moreover their initial average concurrence is diminished by the collisions (Fig.~\ref{fig:fig5} (blue, dotted) and Fig.~\ref{fig:fig5} (red, dashed)) for these network node couples. 

We note that any other ancilla-network interaction (exchange or XX-type) does not result in near-unit concurrence states between network-qubit pairs for this case.

The near-maximally entangled states between the qubits A and C alternate between $\ket{\tilde{\Phi}^+}$ and $\ket{\tilde{\Phi}^-},$ at the peak values similar to the case of linear chain configuration corresponding to Fig.~\ref{fig:fig3a}. Fig.~\ref{fig:fig5} (green) shows a periodic oscillatory structure. The peak value is achieved at sampling index $n=51$ corresponding to the third peak and is $\ket{\tilde{\Phi}^+}$ with a very high fidelity value equal to 0.999. 

Ancilla-reset causes no changes in the entanglement features corresponding to the configuration Fig.~\ref{fig:middle}, therefore Fig.~\ref{fig:fig5} is valid also for the standard collision model resulting in the same Bell-state family at the concurrence maxima as in the continuous-time unitary evolution. 

\section*{Appendix B: different types of entanglement between different qubit pairs due to local x-drive}\label{sec:appB}
\begin{figure}[htb!]
	\centering
	\includegraphics[width=\columnwidth]{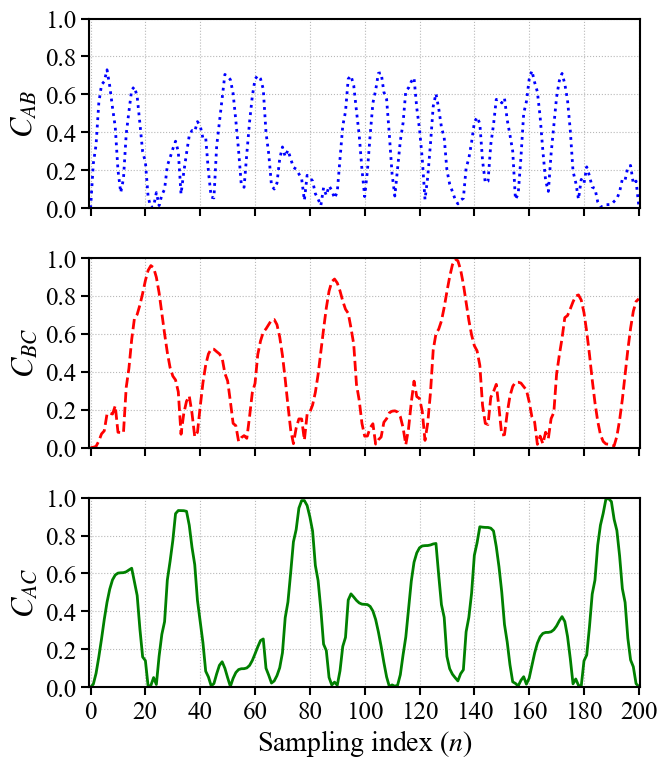}
	\caption{(Color Online) Time evolution of the concurrences $C_\text{AB}(t)$ (blue, dotted), $C_\text{BC}(t)$ (red, dashed), and $C_\text{AC}(t)$ (green) for the open linear chain configuration with the ancilla coupled to the qubit A (Fig.~\ref{fig:edge}). The Hamiltonians are $H_\mathrm{S}=\omega_0[ \left( \sigma^+_\text{A} \sigma^-_\text{B} + \sigma^-_\text{B} \sigma^+_\text{A} \right)+\left( \sigma^+_\text{B} \sigma^-_\text{C} + \sigma^-_\text{C} \sigma^+_\text{B} \right)]/2,$ and $H_\mathrm{int}=\omega\sigma^x_1 \sigma^x_\text{A}$. The parameters are $\omega_0 = 1, \omega = 5, \Delta t = 0.4$. The initial state is  $\ket{+}_1\ket{000}_{\mathrm{ABC}}.$ The ancilla is initialized in the state $\ket{+},$ an eigenstate of $\sigma^x_1.$ In this case, the interaction is equivalent to a coherent local x-drive on qubit A. The continuous-time and fresh-ancilla collision models therefore yield the same network dynamics. }
	\label{fig:fig6}
\end{figure}

We present in Fig.~\ref{fig:fig6} the evolution of concurrence between each pair of qubits for the open linear chain, where the ancilla interacts with qubit A as depicted in Fig.~\ref{fig:edge}. The network qubits interact via nearest neighbor exchange interaction given by
$$ H_\mathrm{S}=\frac{\omega_0}{2}[ \left( \sigma^+_\text{A} \sigma^-_\text{B} + \sigma^-_\text{B} \sigma^+_\text{A} \right)+\left( \sigma^+_\text{B} \sigma^-_\text{C} + \sigma^-_\text{C} \sigma^+_\text{B} \right)].$$ 
The qubit network ABC is initially in the state Eq.~(\ref{eq:initialnetworkstate}) so that the exchange-type interaction between the qubits does not create any correlation with time between the qubits. When the XX-type interaction of the ancilla with qubit A, 
$$H_\mathrm{int}=\omega\sigma^x_1 \sigma^x_\text{A}.$$ 
is turned on, it results in Fig.~\ref{fig:fig6}. The ancilla is initially in the superposition state Eq.~(\ref{eq:initialancillastate}). Since the ancilla is initialized in an eigenstate of $\sigma^x_1,$ the ancilla-network interaction reduces to a local x-drive on qubit A resulting in near-maximal bipartite states between non-adjacent qubits A and C, as well as between qubits B and C, which do not directly interact with the ancilla. Although qubits B and C are not directly driven, the local x-drive applied to qubit A propagates
through the exchange-coupled chain and generates high concurrence between different network-qubit
pairs. Therefore also the continuous-time unitary evolution and collision model calculations yield as expected the same results.

 The qubits B and C are not directly driven, yet they exhibit maximal entanglement with each other at more than one collision step during the simulation (Fig.~\ref{fig:fig6} (red, dashed)). We observe that the concurrence pattern between these qubits is hardly periodic. The maximum value of concurrence is achieved at the index $n=133$ and its value is $C_\text{BC}(n=133)=0.993.$ The achieved state is $\ket{\tilde{\Phi}^-}$ with a fidelity value equal to 0.996.  The peak at $n=22$ is with a lower concurrence value ($C_\text{BC}(n=22)=0.960$) resulting in the state $\ket{\tilde{\Phi}^+}.$ 

The concurrence evolution between qubits A and C (Fig.~\ref{fig:fig6} (green)) is similarly not periodic. These qubits do not directly interact with each other. Still, the interaction with the ancilla leads to high concurrence states between them. 

The peak at $n= 189$ is an entangled state with $C_\text{AC}(n=189)=0.994$ and is one of the triplet Bell states given by 
$$\ket{\Phi^-}=\frac{1}{\sqrt{2}}(\ket{00}-\ket{11}),$$
with a fidelity value equal to 0.989. At the second highest peak at index $n=78,$ a coherent superposition of two different Bell states emerge as
$$\ket{p^-}=\frac{1}{\sqrt{2}}(\ket{\Phi^-}-i\ket{\Psi^-}),$$ 
where $\ket{\Psi^-}=(\ket{01}-\ket{10})/\sqrt{2}$
is the singlet Bell state. The corresponding concurrence value of this state is $C_\text{AC}(n=78)=0.984.$

\newpage
\bibliography{Distributed_ref_rev}

@article{Lieb1972CMP,
	title = {The Finite Group Velocity of Quantum Spin Systems},
	author = {Lieb, Elliott H. and Robinson, Derek W.},
	journal = {Communications in Mathematical Physics},
	volume = {28},
	number = {3},
	pages = {251--257},
	year = {1972},
	publisher = {Springer},
	doi = {10.1007/BF01645779},
	url = {https://doi.org/10.1007/BF01645779}
}

@article{PhysRevLett.97.050401,
	title = {Lieb-Robinson Bounds and the Generation of Correlations and Topological Quantum Order},
	author = {Bravyi, Sergey and Hastings, Matthew B. and Verstraete, Frank},
	journal = {Phys. Rev. Lett.},
	volume = {97},
	issue = {5},
	pages = {050401},
	numpages = {4},
	year = {2006},
	month = {Aug},
	publisher = {American Physical Society},
	doi = {10.1103/PhysRevLett.97.050401},
	url = {https://link.aps.org/doi/10.1103/PhysRevLett.97.050401}
}

@article{PhysRevLett.96.136801,
	title = {Time Dependence of Correlation Functions Following a Quantum Quench},
	author = {Calabrese, Pasquale and Cardy, John},
	journal = {Phys. Rev. Lett.},
	volume = {96},
	issue = {13},
	pages = {136801},
	numpages = {4},
	year = {2006},
	month = {Apr},
	publisher = {American Physical Society},
	doi = {10.1103/PhysRevLett.96.136801},
	url = {https://link.aps.org/doi/10.1103/PhysRevLett.96.136801}
}

@article{PhysRevLett.97.150404,
	title = {General Entanglement Scaling Laws from Time Evolution},
	author = {Eisert, Jens and Osborne, Tobias J.},
	journal = {Phys. Rev. Lett.},
	volume = {97},
	issue = {15},
	pages = {150404},
	numpages = {4},
	year = {2006},
	month = {Oct},
	publisher = {American Physical Society},
	doi = {10.1103/PhysRevLett.97.150404},
	url = {https://link.aps.org/doi/10.1103/PhysRevLett.97.150404}
}

@article{RevModPhys.83.863,
	title = {Colloquium: Nonequilibrium Dynamics of Closed Interacting Quantum Systems},
	author = {Polkovnikov, Anatoli and Sengupta, Krishnendu and Silva, Alessandro and Vengalattore, Mukund},
	journal = {Rev. Mod. Phys.},
	volume = {83},
	issue = {3},
	pages = {863--883},
	numpages = {21},
	year = {2011},
	month = {Aug},
	publisher = {American Physical Society},
	doi = {10.1103/RevModPhys.83.863},
	url = {https://link.aps.org/doi/10.1103/RevModPhys.83.863}
}

@article{Gough2009CMP,
	title = {Quantum Feedback Networks: Hamiltonian Formulation},
	author = {Gough, John and James, Matthew R.},
	journal = {Communications in Mathematical Physics},
	volume = {287},
	number = {3},
	pages = {1109--1132},
	year = {2009},
	publisher = {Springer},
	doi = {10.1007/s00220-008-0698-8},
	url = {https://doi.org/10.1007/s00220-008-0698-8}
}

@article{Gough2009IEEE,
	title = {The Series Product and Its Application to Quantum Feedforward and Feedback Networks},
	author = {Gough, John and James, Matthew R.},
	journal = {IEEE Transactions on Automatic Control},
	volume = {54},
	number = {11},
	pages = {2530--2544},
	year = {2009},
	month = {Nov},
	publisher = {IEEE},
	doi = {10.1109/TAC.2009.2031205},
	url = {https://doi.org/10.1109/TAC.2009.2031205}
}

@article{PhysRevA.78.062104,
	title = {Linear Quantum Feedback Networks},
	author = {Gough, John E. and Gohm, Ralf and Yanagisawa, Masashi},
	journal = {Phys. Rev. A},
	volume = {78},
	issue = {6},
	pages = {062104},
	numpages = {13},
	year = {2008},
	month = {Dec},
	publisher = {American Physical Society},
	doi = {10.1103/PhysRevA.78.062104},
	url = {https://link.aps.org/doi/10.1103/PhysRevA.78.062104}
}

@article{PhysRevA.81.023804,
	title = {Squeezing Components in Linear Quantum Feedback Networks},
	author = {Gough, John and James, Matthew R. and Nurdin, Hendra I.},
	journal = {Phys. Rev. A},
	volume = {81},
	issue = {2},
	pages = {023804},
	numpages = {12},
	year = {2010},
	month = {Feb},
	publisher = {American Physical Society},
	doi = {10.1103/PhysRevA.81.023804},
	url = {https://link.aps.org/doi/10.1103/PhysRevA.81.023804}
}

@article{Gough2016JMP,
	title = {The Stratonovich Formulation of Quantum Feedback Network Rules},
	author = {Gough, John},
	journal = {Journal of Mathematical Physics},
	volume = {57},
	number = {12},
	pages = {122201},
	year = {2016},
	publisher = {AIP Publishing},
	doi = {10.1063/1.4972217},
	url = {https://doi.org/10.1063/1.4972217}
}

@article{PhysRevA.31.3761,
	title = {Input and Output in Damped Quantum Systems: Quantum Stochastic Differential Equations and the Master Equation},
	author = {Gardiner, C. W. and Collett, M. J.},
	journal = {Phys. Rev. A},
	volume = {31},
	issue = {6},
	pages = {3761--3774},
	numpages = {14},
	year = {1985},
	month = {Jun},
	publisher = {American Physical Society},
	doi = {10.1103/PhysRevA.31.3761},
	url = {https://link.aps.org/doi/10.1103/PhysRevA.31.3761}
}

@article{PhysRevLett.70.2269,
	title = {Driving a Quantum System with the Output Field from Another Driven Quantum System},
	author = {Gardiner, C. W.},
	journal = {Phys. Rev. Lett.},
	volume = {70},
	issue = {15},
	pages = {2269--2272},
	numpages = {4},
	year = {1993},
	month = {Apr},
	publisher = {American Physical Society},
	doi = {10.1103/PhysRevLett.70.2269},
	url = {https://link.aps.org/doi/10.1103/PhysRevLett.70.2269}
}

@article{PhysRevLett.70.2273,
	title = {Quantum Trajectory Theory for Cascaded Open Systems},
	author = {Carmichael, H. J.},
	journal = {Phys. Rev. Lett.},
	volume = {70},
	issue = {15},
	pages = {2273--2276},
	numpages = {4},
	year = {1993},
	month = {Apr},
	publisher = {American Physical Society},
	doi = {10.1103/PhysRevLett.70.2273},
	url = {https://link.aps.org/doi/10.1103/PhysRevLett.70.2273}
}

@article{PhysRevA.50.1792,
	title = {Driving Atoms with Light from Other Atoms},
	author = {Gardiner, C. W. and Parkins, A. S.},
	journal = {Phys. Rev. A},
	volume = {50},
	issue = {2},
	pages = {1792--1805},
	numpages = {14},
	year = {1994},
	month = {Aug},
	publisher = {American Physical Society},
	doi = {10.1103/PhysRevA.50.1792},
	url = {https://link.aps.org/doi/10.1103/PhysRevA.50.1792}
}

@article{PhysRevA.87.030101,
	title = {Non-Markovian Master Equations from Piecewise Dynamics},
	author = {Vacchini, Bassano},
	journal = {Phys. Rev. A},
	volume = {87},
	issue = {3},
	pages = {030101},
	numpages = {4},
	year = {2013},
	month = {Mar},
	publisher = {American Physical Society},
	doi = {10.1103/PhysRevA.87.030101},
	url = {https://link.aps.org/doi/10.1103/PhysRevA.87.030101}
}

@article{PhysRevLett.123.180602,
	title = {Collisional Quantum Thermometry},
	author = {Seah, Samuel and Nimmrichter, Stefan and Scarani, Valerio},
	journal = {Phys. Rev. Lett.},
	volume = {123},
	issue = {18},
	pages = {180602},
	numpages = {6},
	year = {2019},
	month = {Oct},
	publisher = {American Physical Society},
	doi = {10.1103/PhysRevLett.123.180602},
	url = {https://link.aps.org/doi/10.1103/PhysRevLett.123.180602}
}

@article{PhysRevA.104.052214,
	title = {Uninformed Bayesian Quantum Thermometry},
	author = {Boeyens, Jan and Seah, Samuel and Nimmrichter, Stefan},
	journal = {Phys. Rev. A},
	volume = {104},
	issue = {5},
	pages = {052214},
	year = {2021},
	month = {Nov},
	publisher = {American Physical Society},
	doi = {10.1103/PhysRevA.104.052214},
	url = {https://link.aps.org/doi/10.1103/PhysRevA.104.052214}
}

@article{PhysRevA.105.042601,
	title = {Bayesian Quantum Thermometry Based on Thermodynamic Length},
	author = {J{\o}rgensen, Mathias R. and Ko{\l}ody{\'n}ski, Jan and Mehboudi, Mohammad and Perarnau-Llobet, Mart{\'i} and Brask, Jonatan B. and Seah, Samuel},
	journal = {Phys. Rev. A},
	volume = {105},
	issue = {4},
	pages = {042601},
	numpages = {11},
	year = {2022},
	month = {Apr},
	publisher = {American Physical Society},
	doi = {10.1103/PhysRevA.105.042601},
	url = {https://link.aps.org/doi/10.1103/PhysRevA.105.042601}
}

@article{Pezzutto2025QST,
	title = {Non-positive Energy Quasidistributions in Coherent Collision Models},
	author = {Pezzutto, Marco and De Chiara, Gabriele and Gherardini, Stefano},
	journal = {Quantum Science and Technology},
	volume = {10},
	issue = {3},
	pages = {035066},
	year = {2025},
	publisher = {IOP Publishing},
	doi = {10.1088/2058-9565/aded2e},
	url = {https://doi.org/10.1088/2058-9565/aded2e}
}

@article{RevModPhys.93.035008,
	title = {Irreversible Entropy Production: From Classical to Quantum},
	author = {Landi, Gabriel T. and Paternostro, Mauro},
	journal = {Rev. Mod. Phys.},
	volume = {93},
	issue = {3},
	pages = {035008},
	numpages = {58},
	year = {2021},
	month = {Sep},
	publisher = {American Physical Society},
	doi = {10.1103/RevModPhys.93.035008},
	url = {https://link.aps.org/doi/10.1103/RevModPhys.93.035008}
}

@article{PhysRevX.7.021003,
	title = {Quantum and Information Thermodynamics: A Unifying Framework Based on Repeated Interactions},
	author = {Strasberg, Philipp and Schaller, Gernot and Brandes, Tobias and Esposito, Massimiliano},
	journal = {Phys. Rev. X},
	volume = {7},
	issue = {2},
	pages = {021003},
	numpages = {24},
	year = {2017},
	month = {Apr},
	publisher = {American Physical Society},
	doi = {10.1103/PhysRevX.7.021003},
	url = {https://link.aps.org/doi/10.1103/PhysRevX.7.021003}
}

@article{PhysRevLett.123.140601,
	title = {Thermodynamics of Weakly Coherent Collisional Models},
	author = {Franklin, Luke and Rodrigues, Samuel and De Chiara, Gabriele and Paternostro, Mauro and Landi, Gabriel T.},
	journal = {Phys. Rev. Lett.},
	volume = {123},
	issue = {14},
	pages = {140601},
	numpages = {6},
	year = {2019},
	month = {Oct},
	publisher = {American Physical Society},
	doi = {10.1103/PhysRevLett.123.140601},
	url = {https://link.aps.org/doi/10.1103/PhysRevLett.123.140601}
}

@article{Corr2025QST,
	title = {Continuous Variable Structured Collision Models},
	author = {Corr, Andrew and Cusumano, Stefano and De Chiara, Gabriele},
	journal = {Quantum Science and Technology},
	volume = {10},
	pages = {015020},
	year = {2025},
	publisher = {IOP Publishing},
	doi = {10.1088/2058-9565/ae0a7b},
	url = {https://doi.org/10.1088/2058-9565/ae0a7b}
}

@article{Cusumano2024NJP,
	title = {Structured Quantum Collision Models: Generating Coherence with Thermal Resources},
	author = {Cusumano, Stefano and De Chiara, Gabriele},
	journal = {New Journal of Physics},
	volume = {26},
	pages = {013018},
	year = {2024},
	month = {Jan},
	publisher = {IOP Publishing},
	doi = {10.1088/1367-2630/ad202a},
	url = {https://doi.org/10.1088/1367-2630/ad202a}
}

@article{PhysRevA.87.040103,
	title = {Collision-Model-Based Approach to Non-Markovian Quantum Dynamics},
	author = {Ciccarello, Francesco and Palma, G. Massimo and Giovannetti, Vittorio},
	journal = {Phys. Rev. A},
	volume = {87},
	issue = {4},
	pages = {040103},
	numpages = {5},
	year = {2013},
	month = {Apr},
	publisher = {American Physical Society},
	doi = {10.1103/PhysRevA.87.040103},
	url = {https://link.aps.org/doi/10.1103/PhysRevA.87.040103}
}

@article{PhysRevA.97.053811,
	title = {Interferometric Modulation of Quantum Cascade Interactions},
	author = {Cusumano, Stefano and Mari, Andrea and Giovannetti, Vittorio},
	journal = {Phys. Rev. A},
	volume = {97},
	issue = {5},
	pages = {053811},
	year = {2018},
	month = {May},
	publisher = {American Physical Society},
	doi = {10.1103/PhysRevA.97.053811},
	url = {https://link.aps.org/doi/10.1103/PhysRevA.97.053811}
}

@article{PhysRevA.95.053838,
	title = {Interferometric Quantum Cascade Systems},
	author = {Cusumano, Stefano and Mari, Andrea and Giovannetti, Vittorio},
	journal = {Phys. Rev. A},
	volume = {95},
	issue = {5},
	pages = {053838},
	numpages = {12},
	year = {2017},
	month = {May},
	publisher = {American Physical Society},
	doi = {10.1103/PhysRevA.95.053838},
	url = {https://link.aps.org/doi/10.1103/PhysRevA.95.053838}
}

@article{CICCARELLO20221,
	title = {Quantum collision models: Open system dynamics from repeated interactions},
	journal = {Physics Reports},
	volume = {954},
	pages = {1-70},
	year = {2022},
	issn = {0370-1573},
	doi = {https://doi.org/10.1016/j.physrep.2022.01.001},
	url = {https://www.sciencedirect.com/science/article/pii/S0370157322000035},
	author = {Francesco Ciccarello and Salvatore Lorenzo and Vittorio Giovannetti and G. Massimo Palma}
}

@article{Horodecki2009,
  title={Quantum entanglement},
  author={Horodecki, Ryszard and Horodecki, Pawel and Horodecki, Michal and Horodecki, Karol},
  journal={Rev. Mod. Phys.},
  volume={81},
  pages={865--942},
  year={2009},
  doi={10.1103/RevModPhys.81.865}
}

@article{Pezze2018,
  title={Quantum metrology with nonclassical states of atomic ensembles},
  author={Pezze, Luca and Smerzi, Augusto and Oberthaler, Markus K. and Schmied, Roman and Treutlein, Philipp},
  journal={Rev. Mod. Phys.},
  volume={90},
  number={3},
  pages={035005},
  year={2018},
  doi={10.1103/RevModPhys.90.035005}
}

@article{Giovannetti2006,
  title={Quantum Metrology},
  author={Giovannetti, V. and Lloyd, S. and Maccone, L.},
  journal={Phys. Rev. Lett.},
  volume={96},
  pages={010401},
  year={2006},
  doi={10.1103/PhysRevLett.96.010401}
}

@article{Toth2014,
  title={Quantum metrology from a quantum information science perspective},
  author={T{\'o}th, G{\'e}za and Apellaniz, Iagoba},
  journal={J. Phys. A: Math. Theor.},
  volume={47},
  number={42},
  pages={424006},
  year={2014},
  doi={10.1088/1751-8113/47/42/424006}
}

@article{Monz2011,
  title={14-Qubit Entanglement: Creation and Coherence},
  author = {Monz, Thomas and Schindler, Philipp and Barreiro, Julio T. and Chwalla, Michael and Nigg, Daniel and Coish, William A. and Harlander, Maximilian and H\"ansel, Wolfgang and Hennrich, Markus and Blatt, Rainer},
  journal={Phys. Rev. Lett.},
  volume={106},
  pages={130506},
  year={2011},
  doi={10.1103/PhysRevLett.106.130506}
}

@article{Song2019,
  title={Generation of multicomponent atomic Schr{\"o}dinger cat states of up to 20 qubits},
  author = {Chao Song and Kai Xu and Hekang Li and Yu-Ran Zhang and
  Xu Zhang and Wuxin Liu and Qiujiang Guo and Zhen Wang and
  Wenhui Ren and Jie Hao and Hui Feng and Heng Fan and
  Dongning Zheng and Da-Wei Wang and H. Wang and Shi-Yao Zhu},
  title = {Generation of Multicomponent Atomic Schrödinger Cat States of up to 20 Qubits},
  journal = {Science},
  volume = {365},
  number = {6453},
  pages = {574--577},
  year = {2019},
  doi = {10.1126/science.aay0600}
}

@article{Zwerger2012,
  title={Measurement-based quantum repeaters},
  author = {Zwerger, M. and D\"ur, W. and Briegel, H. J.},
  journal={Phys. Rev. A},
  volume={85},
  pages={062326},
  year={2012},
  doi={10.1103/PhysRevA.85.062326}
}

@article{Zukowski1993,
  title={“Event-ready-detectors” Bell experiment via entanglement swapping},
  author={\.{Z}ukowski, Marek and Zeilinger, Anton and Horne, Michael A. and Ekert, Artur K.},
  journal={Phys. Rev. Lett.},
  volume={71},
  pages={4287},
  year={1993},
  doi={10.1103/PhysRevLett.71.4287}
}

@article{Kimble2008,
  title={The quantum internet},
  author={Kimble, H. J.},
  journal={Nature},
  volume={453},
  number={7198},
  pages={1023--1030},
  year={2008},
  doi={10.1038/nature07127}
}

@article{Ciccarello2017,
  title={Collision models in open system dynamics: A versatile tool for deeper insights?},
  author={Ciccarello, Francesco},
  journal={Quantum Meas. Quantum Metrol.},
  volume={4},
  pages={53--63},
  year={2017},
  doi={10.1515/qmetro-2017-0007}
}

@article{Scarani2002,
  title={Thermalizing quantum machines: Dissipation and entanglement},
  author={Scarani, Valerio and Ziman, Mario and Štelmachovič, Peter and Gisin, Nicolas and Bužek, Vladimír},
  journal={Phys. Rev. Lett.},
  volume={88},
  pages={097905},
  year={2002},
  doi={10.1103/PhysRevLett.88.097905}
}

@article{Karevski2009,
  title={Quantum nonequilibrium steady states induced by repeated interactions},
  author={Karevski, D. and Platini, T.},
  journal={Phys. Rev. Lett.},
  volume={102},
  pages={207207},
  year={2009},
  doi={10.1103/PhysRevLett.102.207207}
}

@article{McCloskey2014,
  title={Non-Markovianity and System-Environment Correlations in a Microscopic Collision Model},
  author={McCloskey, R. and Paternostro, M.},
  journal={Phys. Rev. A},
  volume={89},
  pages={052120},
  year={2014},
  doi={10.1103/PhysRevA.89.052120}
}

@article{Lorenzo2017,
  title={Non-Markovian dynamics from collision models with interancilla correlations},
  author={Lorenzo, S. and Ciccarello, F. and Palma, G. M.},
  journal={Phys. Rev. A},
  volume={96},
  pages={032107},
  year={2017},
  doi={10.1103/PhysRevA.96.032107}
}

@article{Ziman2005,
  title={Entanglement induced by a single-mode heat bath},
  author={Ziman, M. and Bužek, V.},
  journal={Phys. Rev. A},
  volume={72},
  pages={022110},
  year={2005},
  doi={10.1103/PhysRevA.72.022110}
}

@article{Giovannetti2012,
  title={Environment-assisted entanglement distribution},
  author={Giovannetti, V. and Palma, G. M. and Sciarrino, F.},
  journal={Phys. Rev. A},
  volume={86},
  pages={012302},
  year={2012},
  doi={10.1103/PhysRevA.86.012302}
}

@article{Yosifov2025,
  title={Dissipation-induced quantum homogenization for temporal information processing},
  author={Yosifov, A. and Iyer, A. and Vedral, V.},
  journal={Phys. Rev. A},
  volume={111},
  pages={012622},
  year={2025}
}

@article{Yosifov2024,
  title={Quantum homogenization as a quantum steady-state protocol on noisy intermediate-scale quantum hardware},
  author={Yosifov, A. and Iyer, A. and Ebler, D. and Vedral, V.},
  journal={Phys. Rev. A},
  volume={109},
  pages={032624},
  year={2024}
}

@article{Karpat2025,
  title={Transient dynamics and homogenization in incoherent collision models},
  author={Karpat, Göktuğ and Çakmak, Barış},
  journal={Entropy},
  volume={27},
  number={2},
  pages={206},
  year={2025},
  doi={10.3390/e27020206}
}

@article{Bell1964,
	title = {On the Einstein Podolsky Rosen paradox},
	author = {Bell, J. S.},
	journal = {Physics Physique Fizika},
	volume = {1},
	issue = {3},
	pages = {195--200},
	numpages = {6},
	year = {1964},
	month = {Nov},
	publisher = {American Physical Society},
	doi = {10.1103/PhysicsPhysiqueFizika.1.195},
	url = {https://link.aps.org/doi/10.1103/PhysicsPhysiqueFizika.1.195}
}

@article{Bennett1993,
	title = {Teleporting an unknown quantum state via dual classical and Einstein-Podolsky-Rosen channels},
	author = {Bennett, Charles H. and Brassard, Gilles and Crepeau, Claude and Jozsa, Richard and Peres, Asher and Wootters, William K.},
	journal = {Phys. Rev. Lett.},
	volume = {70},
	issue = {13},
	pages = {1895--1899},
	numpages = {0},
	year = {1993},
	month = {Mar},
	publisher = {American Physical Society},
	doi = {10.1103/PhysRevLett.70.1895},
	url = {https://link.aps.org/doi/10.1103/PhysRevLett.70.1895}
}

@article{Wooters1998,
	title = {Entanglement of Formation of an Arbitrary State of Two Qubits},
	author = {Wootters, William K.},
	journal = {Phys. Rev. Lett.},
	volume = {80},
	issue = {10},
	pages = {2245--2248},
	numpages = {0},
	year = {1998},
	month = {Mar},
	publisher = {American Physical Society},
	doi = {10.1103/PhysRevLett.80.2245},
	url = {https://link.aps.org/doi/10.1103/PhysRevLett.80.2245}
}

@article{Bennett1996,
	title = {Mixed-state entanglement and quantum error correction},
	author = {Bennett, Charles H. and DiVincenzo, David P. and Smolin, John A. and Wootters, William K.},
	journal = {Phys. Rev. A},
	volume = {54},
	issue = {5},
	pages = {3824--3851},
	numpages = {0},
	year = {1996},
	month = {Nov},
	publisher = {American Physical Society},
	doi = {10.1103/PhysRevA.54.3824},
	url = {https://link.aps.org/doi/10.1103/PhysRevA.54.3824}
}

@article{PhysRevLett.78.5022,
	title = {Entanglement of a Pair of Quantum Bits},
	author = {Hill, Sam A. and Wootters, William K.},
	journal = {Phys. Rev. Lett.},
	volume = {78},
	issue = {26},
	pages = {5022--5025},
	numpages = {0},
	year = {1997},
	month = {Jun},
	publisher = {American Physical Society},
	doi = {10.1103/PhysRevLett.78.5022},
	url = {https://link.aps.org/doi/10.1103/PhysRevLett.78.5022}
}

@article{Wootters2001,
	author = {Wootters, William K.},
	title = {Entanglement of formation and concurrence},
	year = {2001},
	issue_date = {January 2001},
	publisher = {Rinton Press, Incorporated},
	address = {Paramus, NJ},
	volume = {1},
	number = {1},
	issn = {1533-7146},
	journal = {Quantum Info. Comput.},
	month = jan,
	pages = {27–44},
	numpages = {18}
}

\end{document}